# Implicitly Learned Neural Phase Functions for Point Spread Function Engineering

Aleksey Valouev*
*Boston University*
Boston, USA
alekseyv@bu.edu

***Abstract*— Point spread function (PSF) engineering is vital for precisely controlling the focus of light in computational imaging, with applications in neural imaging, fluorescence microscopy, and biophotonics. The PSF is derived from the magnitude of the Fourier transform of a phase function, making the construction of the phase function given the PSF (PSF engineering) an ill-posed inverse problem. Traditional PSF engineering methods rely on physical basis functions, limiting their ability to generalize across the range of PSFs required for imaging tasks. We introduce a novel approach leveraging implicit neural representations that overcome the limitations of pixel-wise optimization methods. Our approach achieves a median MSSIM of 0.8162 and a mean MSSIM of 0.5634, compared to a median MSSIM of 0.0 and a mean MSSIM of 0.1841 with pixel-wise optimization when learning randomly generated phase functions. Our approach also achieves a median PSNR of 10.38 dB and a mean PSNR of 8.672 dB, compared to a median PSNR of 6.653 dB and a mean PSNR of 6.660 dB with pixel-wise optimization for this task.**

***Keywords*— *computational imaging, point spread function engineering, implicit neural representation***

## I. INTRODUCTION

Recently, the field of computational imaging has experienced rapid growth, with significant advancements in areas such as super-resolution microscopy and particle tracking [1], imaging of in vivo neuronal populations [2], and biophotonics [3]. A central challenge in these applications is to design point spread functions (PSFs) to precisely focus light, via precise and robust PSF engineering.

## II. PSF ENGINEERING

### *A. Defining the PSF*

The PSF of an optical system is determined by the magnitude of the Fourier transform of a phase function situated within the system's Fourier plane. Figure 1 illustrates an optical system where a spatial light modulator (SLM) projects a specified phase function. Equation 3 outlines the process for calculating a PSF from the phase function. Note that the operation $|\mathcal{F}(\cdot)|^2$ is non-invertible, and therefore retrieving a phase function from a given PSF is an ill-posed inverse problem.

### *B. Current Approaches*

Several approaches have been proposed to address the ill-posed inverse problem, including genetic algorithms for optimizing the coefficients of aspheric functions [2], coefficient-solving algorithms (CSAs) for Zernike basis functions [4], and pixel-wise, gradient-based optimization methods (Fig. 2) [1]. However, methods that rely on physical basis functions are ineffective due to the incomplete and non-orthogonal nature of physical bases, which limits their ability to represent the full space of possible phase functions.

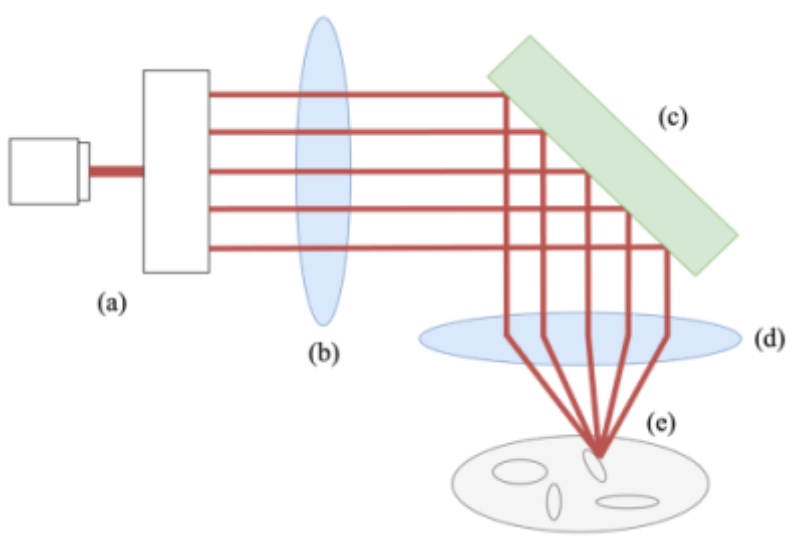

Fig. 1. Simplified model of an optical system with a SLM. Laser emits a light beam (a) which passes through a Fourier transform lens (b). The phase of the light is modulated when it bounces off the SLM (c) and passes through inverse Fourier transform lens (d), which focuses the light (e) and forms the PSF.

Currently, the most effective of these approaches is pixel-wise optimization, where gradient descent is performed on discrete pixels [1]. However, this approach has significant limitations: the resolution of the resulting phase function is determined by the resolution of the pixel grid, meaning resampling the phase function to be displayed at different resolutions is not possible.

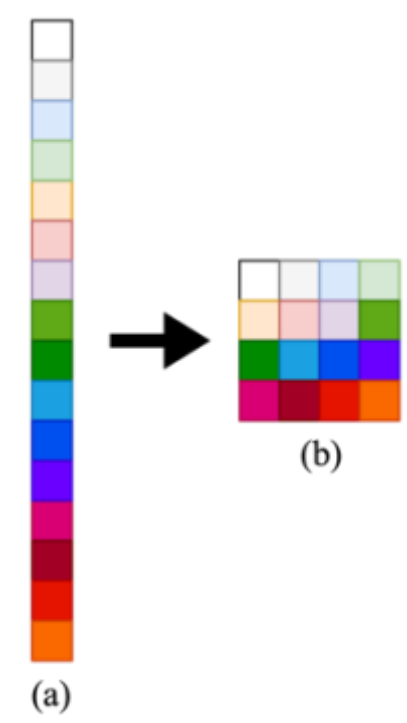

Fig. 2. Pixelwise optimization. Each pixel is represented as a trainable parameter (a). Phase functions is formed by resizing the optimized vector (b).

### *C. Implicit Neural Representation*

A recent breakthrough for solving ill-posed inverse problems is the development of implicit neural representations [5]. Initially applied in novel view synthesis and 3D scene representation, this method has been successfully used in computational imaging, particularly for phase retrieval in adaptive optics [6]. Implicit neural representations offer an advantage over pixel-wise approaches since they produce smooth and differentiable phase functions that are easier to manufacture. Given these advantages, we expect that implicit neural representations are well suited for the task of PSF engineering.

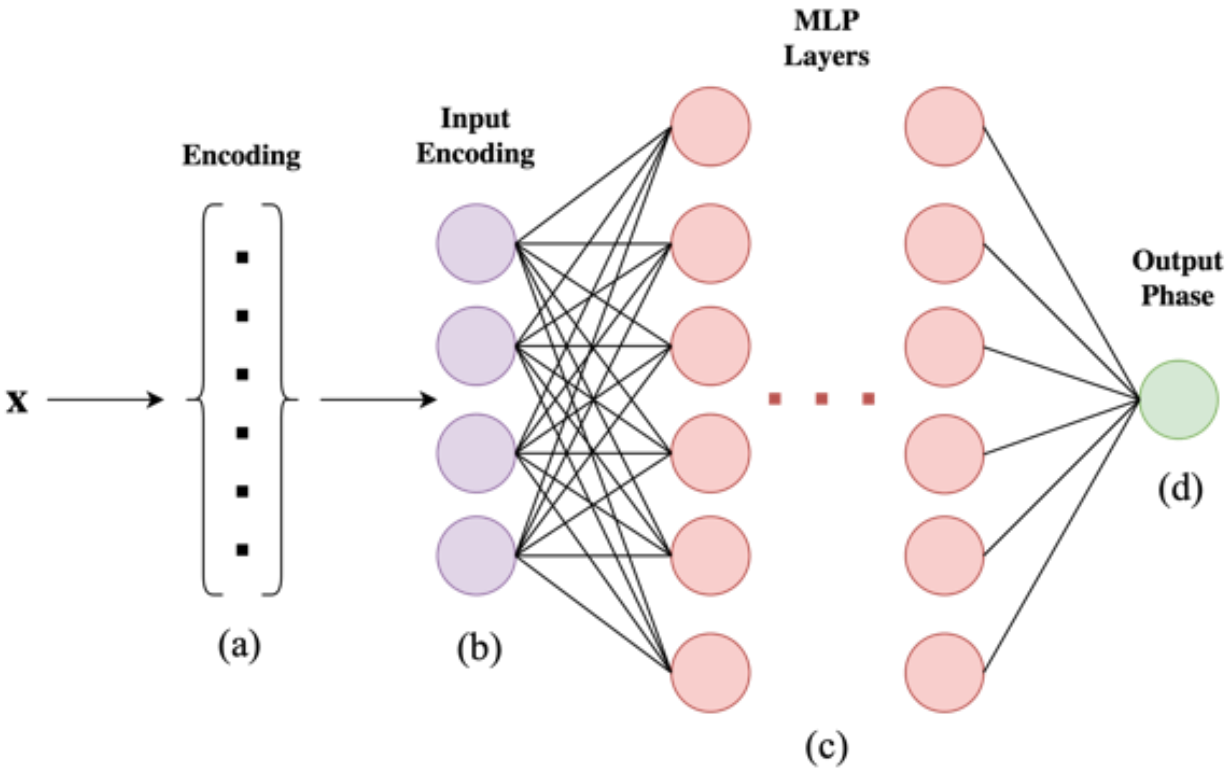

Fig. 3. Implicit neural representation architecture. Each (x, y) input is encoded using radial encoding scheme (a). The resulting vector is used as the input for the first fully-connected layer (b). The subsequent MLP layers are formed using a fully-connected layer followed by a Leaky ReLU activation (c). The final fully-connected layer outputs a single value, the phase at the point (x, y) (d).

## III. Methods

We define our implicit neural representation (Fig. 3) using the function:

$$F_{\Theta}(x, y) \tag{1}$$

To obtain the phase function, we sample points $(x_i, y_j)$ on a grid given by:

$$x_i \in \{x_0, \dots x_{i_{\max}}\},\ y_j \in \{y_0, \dots y_{j_{\max}}\}, z_k \in \{z_0, \dots z_{k_{\max}}\} \tag{2}$$

Our goal is to learn a PSF using the following predictor:

$$\widehat{PSF}(x_i, y_j, z_k) = |\mathcal{F}(e^{iF_{\Theta}(x_i, y_j)} D(x_i, y_j, z_k)|^2 \tag{3}$$

with $\mathcal{F}$ being the Fourier transform operator and $D$ being the defocus kernel of the optical system [7] (Fig. 4).

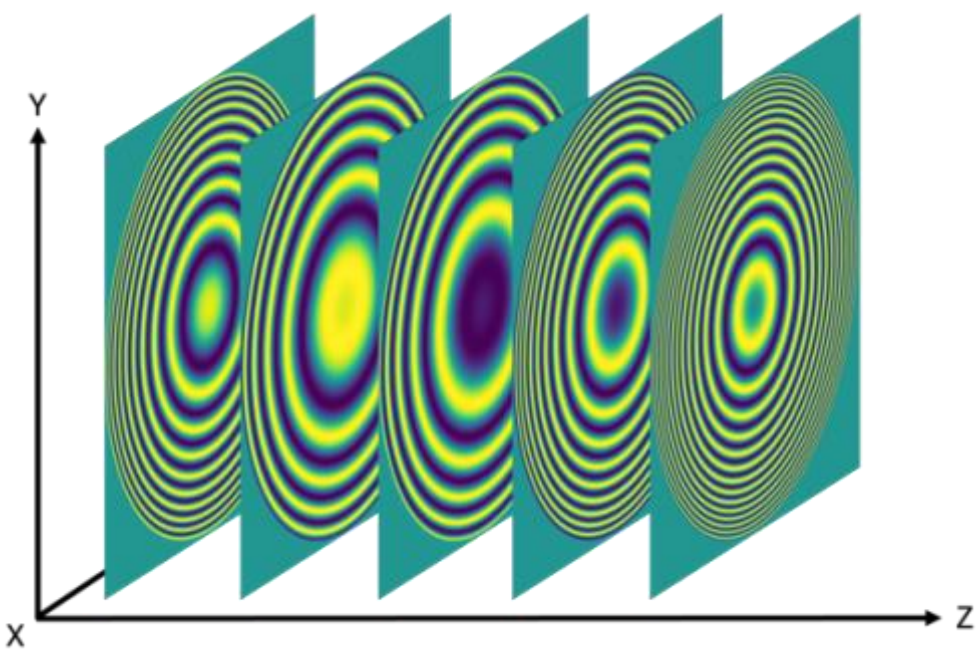

Fig. 4. Real component of the defocus kernel.

We train the predictor using a mean-squared error loss:

$$\mathcal{L} = \frac{1}{i_{\max} j_{\max} k_{\max}} \sum_{i=0}^{i_{max}} \sum_{j=0}^{j_{max}} \sum_{k=0}^{k_{max}} \left( \widehat{PSF}(x_i, y_j, z_k) - PSF(x_i, y_j, z_k) \right)^2 \tag{4}$$

To overcome spectral bias [8, 9] we use a radial encoding scheme [10]. We define a set of rotations for input (x, y):

$$\begin{pmatrix} x'_K \\ y'_K \end{pmatrix} = \begin{bmatrix} cos(\theta_K) & -\sin(\theta_K) \\ \sin(\theta_K) & \cos(\theta_K) \end{bmatrix} \begin{pmatrix} x \\ y \end{pmatrix} \tag{5}$$

$$\theta_K = \frac{2\pi K}{K_{max}+1}, K \in \{0, 1, 2, \dots K_{max}\} \tag{6}$$

Taking $L$ as the encoding length, the $K$ th encoding sequence is then defined with:

$$\gamma_K(x, y) = \begin{pmatrix} \cos(2^0 \pi x'_K) \\ \cos(2^0 \pi y'_K) \\ \sin(2^0 \pi x'_K) \\ \sin(2^0 \pi y'_K) \\ \vdots \\ \cos(2^{L-1} \pi x'_K) \\ \cos(2^{L-1} \pi y'_K) \\ \sin(2^{L-1} \pi x'_K) \\ \sin(2^{L-1} \pi y'_K) \end{pmatrix} \tag{7}$$

Our full encoding scheme concatenates all $\gamma_K$ from $K = 0$ to $K_{max}$. The full structure of our MLP is described in Figure 3, and our training scheme is described in Figure 5. We use 3 hidden layers of size 256, and a Leaky ReLU activation function. We use a radial encoding with $K_{max} = 10$ and $L = 40$. During training, we use a cosine annealing learning rate scheduler [11] to decay the learning rate from 0.0005 to 0.0001 over 10,000 epochs.

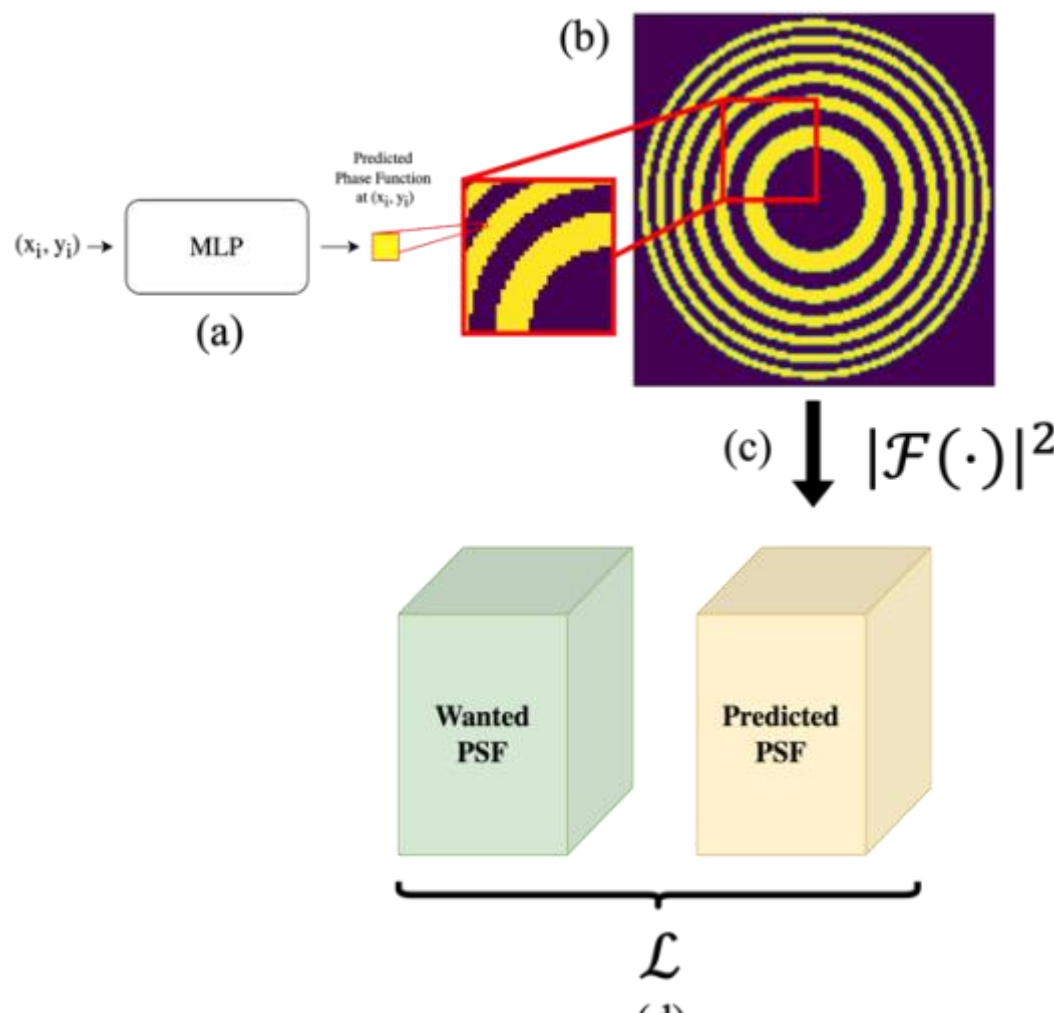

Fig. 5. Training architecture for models. Phase at input $(x_i, y_j)$ coordinate is found using forward pass (a). Process is repeated at each pixel on phase function (b). Fourier transform is applied and PSF is obtained (c). Loss is calculated based on the predicted PSF and the target PSF, and gradient descent is performed (d).

## IV. Results

To evaluate our model's performance, we used the Multiscale Structural Similarity (MSSIM) [12] and peak signal-to-noise ratio (PSNR) as metrics. We constructed a set of 40 randomly initialized phase functions, with values at each pixel coordinate sampled from a uniform distribution $U(0,1)$. Both the pixel-wise network and the implicit neural representation were trained on PSFs generated from these randomly initialized pupils. The implicit neural representation outperformed the pixel-wise method across all metrics, as shown in Table 1. Additionally, we assessed the performance of both models on phase functions constructed from the superposition of ten randomly weighted Zernike basis functions. Our results show that the performance of the two models is comparable for this task, however the pixelwise model sometimes produces visual artifacts (Table 2, Fig. 6).

TABLE I. RANDOM PUPILS RESULTS

**Evaluation Metrics for Models Trained on Random Pupils (n=40)**

| | *Implicit Neural Representation* | *Pixel-Wise Method* |
|---|---|---|
| Median MSSIM | **0.8162** | 0.0 |
| Mean MSSIM | **0.5634** | 0.1841 |
| STD MSSIM | 0.4202 | 0.2070 |
| Median PSNR (dB) | **10.38** | 6.653 |
| Mean PSNR (dB) | **8.672** | 6.660 |
| STD PSNR (dB) | 7.038 | 0.0534 |

TABLE II. ZERNIKE PUPILS RESULTS

**Evaluation Metrics for Models Trained on Zernike Pupils (n=40)**

| | *Implicit Neural Representation* | *Pixel-Wise Method* |
|---|---|---|
| Median MSSIM | 0.8525 | **0.8558** |
| Mean MSSIM | 0.7138 | **0.7295** |
| STD MSSIM | 0.3263 | 0.3225 |
| Median PSNR (dB) | **14.59** | 12.91 |
| Mean PSNR (dB) | **16.54** | 14.00 |
| STD PSNR (dB) | 8.321 | 5.788 |

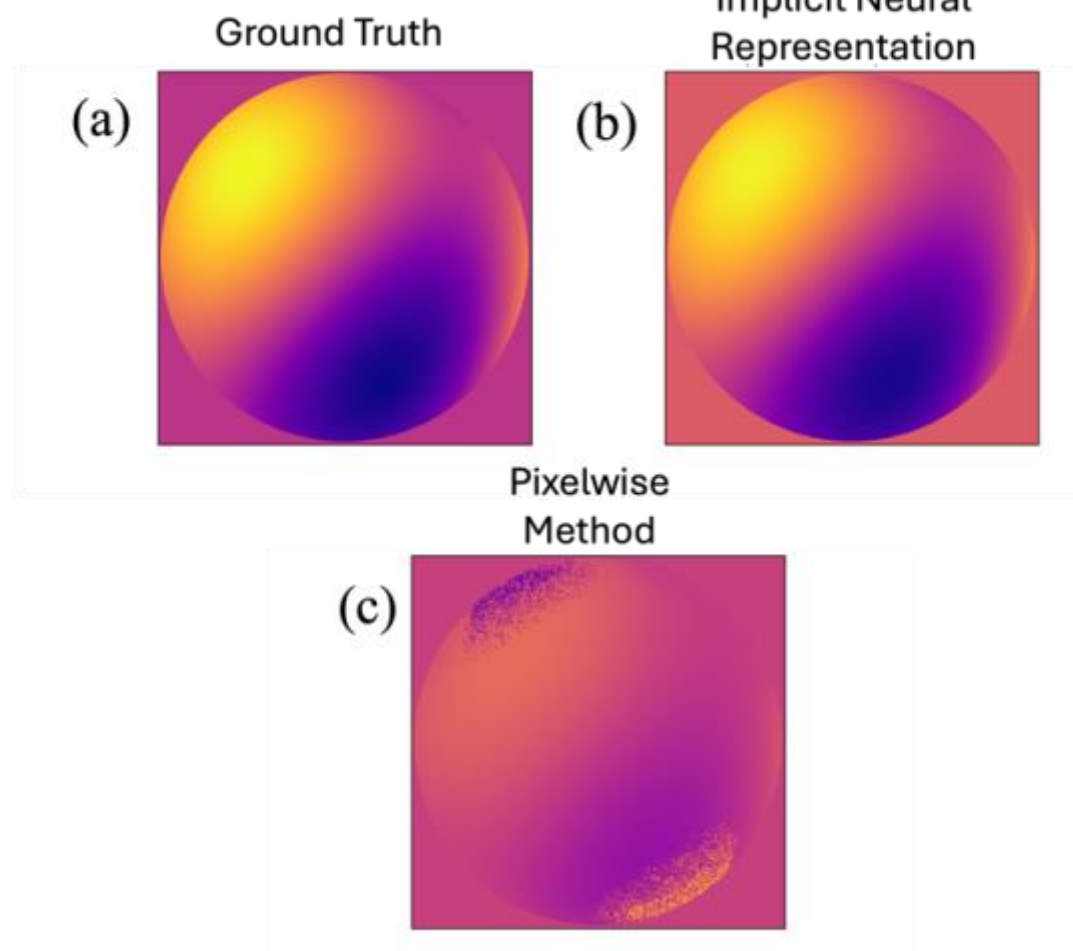

Fig. 6. Results for a randomly initialized Zernike pupil. Implicit neural representation (b) is closer to ground truth (a) than pixel-wise method (c). Pixelwise pupil has noisy pixels and artifacts near the edges.

We also evaluated the ability of both models to learn highly oscillatory phase functions by training on an extended depth-of-field (EdoF) PSF. We find that the implicit neural representation is not able to learn high-frequencies as well as the pixel-wise method (Fig. 7). However, the phase function learned by the pixelwise method was much "noisier" than the implicit neural representation.

Overall, our implicit neural representation outperformed the pixel-wise method in learning a variety of randomly initialized pupils, as measured by MSSIM and PSNR. However, the pixel-wise method outperforms the implicit neural representation when learning phase functions with more high frequency components present.

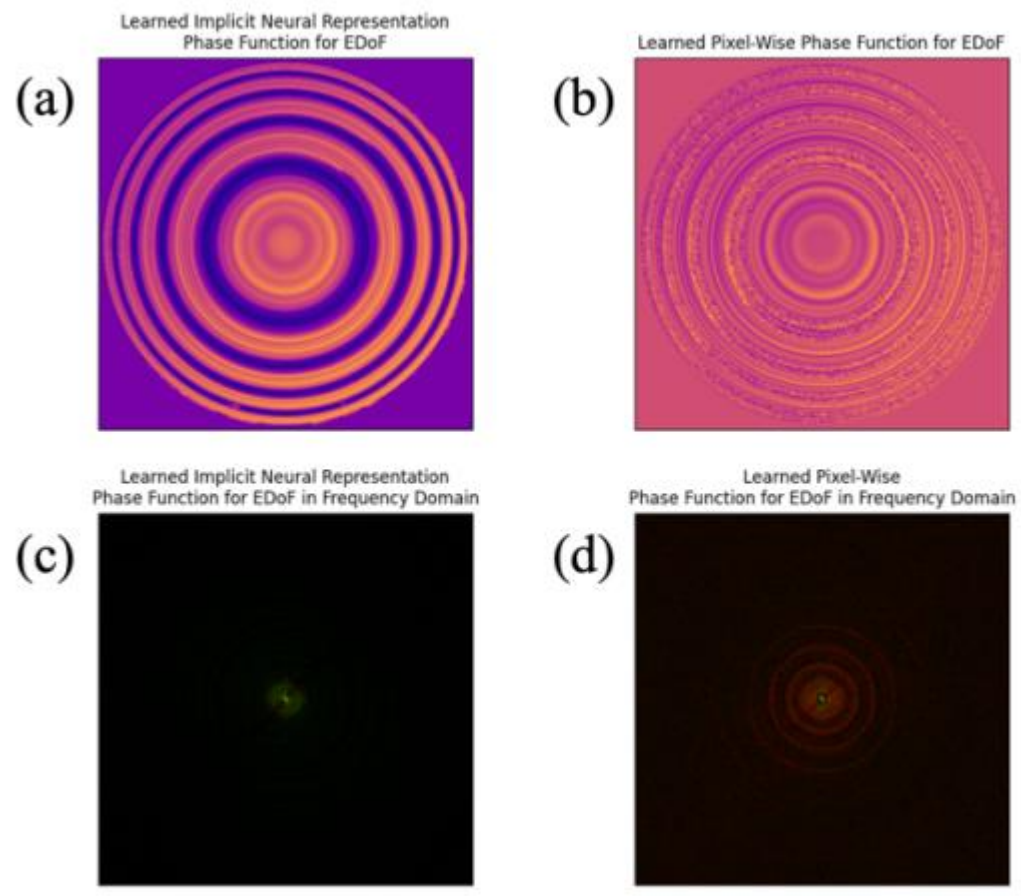

Fig. 7. Phase learned by implicit neural representation (a) oscillates less than phase learned by pixel-wise method (b) for an EDoF PSF. Analysis of the phase function in the frequency domain (low frequency components at origin) shows that pixel-wise phase function (d) learns higher frequency components better than implicit neural representation (c). RGB values in (c) and (d) scaled using sine function for visibility.

## V. DISCUSSION

In this work, we demonstrate that implicit neural representations significantly outperform the current state-of-the-art pixel-wise method for PSF engineering when learning arbitrary PSFs. However, the pixel-wise method outperforms the implicit neural representation when learning phase functions that have many high frequency components present. This result presents a key difference between the pixel-wise method and the implicit neural representation: the pixel-wise method is better at learning highly oscillatory phase functions, while the implicit neural representation is better at learning continuous and smooth phase functions. Overall, the implicit neural representation shows significant potential due to increased performance for learning a wider variety of phase functions.

Beyond performance improvements, the neural network-based nature of implicit representations enables the use of advanced training techniques, such as transfer learning [13] and sinusoidal activation functions [14], which can further enhance model performance. Transfer learning offers the potential to significantly reduce training time, enabling real-time PSF engineering and allowing for rapid adaptation of the PSF to focus light at different points within a medium.

This research presents several key advancements in the field of computational imaging. First, our implicit neural representation approach is the best performing method for pupil engineering to date, outperforming the pixel-wise approach across a diverse set of pupils. Second, the ability to integrate the implicit neural representation into an end-to-end optimization framework allows the model to be trained to optimize imaging systems for specific tasks. Overall, the use of implicit neural representations for PSF engineering has significant future implications for deep optics and computational imaging because it eliminates the need for a physical basis for PSF engineering and pioneers a "plug-and-play" approach to PSF engineering.